\documentclass[aps,pra,numerical,superscriptaddress,showpacs,amsmath,amssymb,
floatfix,
reprint,
]{revtex4-1}
\usepackage{graphicx}
\usepackage{dcolumn}
\usepackage[english]{babel}
\usepackage[noload=abbr]{siunitx}
\usepackage{nicefrac}
\usepackage{tikz}
\usetikzlibrary{shapes}

\newunit{\dBm}{dBm}

\begin{document}
\title{The interplay between thermal Rydberg gases and plasmas}
\author{Daniel Weller}
\affiliation{5.~Physikalisches Institut and Center for Integrated Quantum Science and Technology, 
University of Stuttgart, 
Pfaffenwaldring 57,
70569 Stuttgart, 
Germany}
\author{James P.\,Shaffer}	
\affiliation{Quantum Valley Ideas Laboratories,
485 Wes Graham Way,
Waterloo, ON N2L 0A7,
Canada}
\author{Tilman Pfau}
\affiliation{5.~Physikalisches Institut and Center for Integrated Quantum Science and Technology, 
University of Stuttgart, 
Pfaffenwaldring 57,
70569 Stuttgart, 
Germany}
\author{Robert L\"ow}
\affiliation{5.~Physikalisches Institut and Center for Integrated Quantum Science and Technology, 
University of Stuttgart, 
Pfaffenwaldring 57,
70569 Stuttgart, 
Germany}
\author{Harald K\"ubler}
\email{h.kuebler@physik.uni-stuttgart.de}
\homepage{\newline http://www.pi5.uni-stuttgart.de}
\affiliation{5.~Physikalisches Institut and Center for Integrated Quantum Science and Technology, 
University of Stuttgart, 
Pfaffenwaldring 57,
70569 Stuttgart, 
Germany}

\date{\today}

\begin{abstract}
We investigate the phenomenon of bistability
in a thermal gas of cesium atoms
excited to Rydberg states.
We present both measurements and a numerical model of the phenomena based on collisions.
By directly measuring the plasma frequency,
we show that the origin of the bistable behavior
lies in the creation of a plasma
formed by ionized Rydberg atoms.
Recombination of ions and electrons manifests
as fluorescence which
allows us to characterize the plasma properties
and study the transient dynamics of the hysteresis that occurs.
We determine scaling parameters for
the point of plasma formation,
and verify our numerical model by comparing measured and simulated spectra.
These measurements yield a detailed microscopic picture
of ionization and avalanche processes occurring in thermal Rydberg gases.
From this set of measurements,
we conclude that plasma formation is a fundamental ingredient in the optical bistability taking place in thermal Rydberg gases
and imposes a limit on usable Rydberg densities for many applications.
\end{abstract}
\pacs{42.65.Pc, 32.80.Rm, 34.50.Fa, 34.20.Cf}
%

\maketitle
\section{Introduction} 
\label{sec:introduction}
Highly excited atoms are well-known for their extraordinary properties,
such as high sensitivity to electric fields
and strong interactions between Rydberg atoms
effective in the micrometer range.
These properties can be fine-tuned
by choosing different Rydberg states.
The strong long-range interactions between Rydberg atoms
has motivated studies of a variety of collective phenomena
and applications with strongly correlated atomic clouds.
Among some of the most prominent in the field of ultracold atoms are
quantum gates~\citep{saffman2005analysis,jones2007fast},
quantum phase transitions~\citep{low2009universal,schauss2014dynamical}, 
optical non-linearities on the single photon level~\citep{dudin2012strongly,peyronel2012quantum,maxwell2013storage,tiarks2014single,gorniaczyk2014single}, 
beyond two-body interactions~\citep{faoro2015borromean},
excitation transfer~\citep{gunter2013observing,barredo2015coherent},
aggregation of excitations~\citep{schempp2014full,malossi2014full,urvoy2015strongly}
and ultralong-range molecules~\citep{bendkowsky2009observation,Overstreet2009}.
The study of Rydberg atoms in thermal vapors
enriches the spectrum of research topics
as these vapors can provide much larger atom numbers
and higher densities compared to ultracold gases.
Besides this,
the technical overhead for hot vapor spectroscopy
is much smaller
and it does not require preparation steps
such as laser and evaporative cooling.
The possibility to work at high bandwidths 
and fast timescales
brings this technology closer to real world applications.
Examples of these areas of focus are
electric field and terahertz sensing~\citep{sedlacek2013atom,wade2018terahertz}
and single photon sources~\citep{ripka2018room}.
One of the topics that has generated interest in hot Rydberg atom physics
is the phenomenon of
optical bistabilities~\citep{carr2013nonequilibrium,marcuzzi2014universal,
vsibalic2015driven,de2016intrinsic,ding2016non,weller2016charge}.
The mechanism that is responsible
for the observation of bistability in Rydberg gases has been controversial
as charged particles and long-range Rydberg atom interactions can,
in principle,
both lead to bistable behavior.
In our previous publication~\citep{weller2016charge}
we presented experimental evidence that the underlying mechanism
for the optical bistability is given by the presence of charges in the vapor.
However,
a minute picture remains to be worked out.
In this article,
we show that the appearance of optical bistability in a thermal Rydberg vapor
is caused by a plasma that is formed due to inelastic, ionizing collisions
between Rydberg atoms and ground state atoms which generate charged particles.
The electrons produced by the ionization of the Rydberg atoms
become an additional source for collisions with Rydberg atoms
leading to even more charges in the system.
Eventually, when sufficiently many Rydberg atoms are ionized,
a steady state plasma is formed.
We explore the relation between the Rydberg density,
the plasma charge density and the plasma frequency
by directly measuring the interaction of the plasma
with a radio-frequency field.
We introduce a model
for the steady state of the system
based on the excitation of Rydberg states 
and their subsequent ionization due to collisions.
With this model we are able to determine two scaling parameter
(one for each scan direction of the excitation lasers)
such that the point of plasma formation
is described by a simple linear function.
The theoretical model is compared to 
experimental
spectra and excellent agreement is found.
We conclude that 
the underlying mechanism for the
optical bistability in thermal Rydberg gases is
dominated by the formation of a plasma,
which acts back on the Rydberg excitation
via the Stark effect.
The interplay between Rydberg density and the presence of a plasma
has important implications on every thermal vapor cell experiment
that involves Rydberg atoms
when laser intensities are outside the weak probe regime.
The presence of charged particles in the vapor
alters the value of the electric dipole matrix elements of the atoms
and allows for the admixture of dipole-forbidden transitions.
This directly affects the pair-potentials of Rydberg-Rydberg interactions
and therefore influences any blockade/anti-blockade physics
as well as Rydberg dressing.
As the electric field is not homogeneous but given by a distribution,
the field gradients lead to shorter coherence times
and time-varying interactions.
Especially continuous beam experiments suffer from this limitation,
but also pulsed experiments are imposed by an upper boundary
with respect to the repetition rate and pulse duration.
Understanding the microscopic mechanisms
that are responsible for ion production
and its corresponding electric field distribution
in thermal Rydberg vapor is an important component
for future experiments and applications.
The paper is organized in 6 sections.
In Sec.~\ref{sec:experiment} 
we describe the experimental setup and conditions
under which we performed the measurements.
We qualitatively describe the microscopic model
that captures the underlying mechanisms responsible
for the observed phenomenon of optical bistability in Sec.~\ref{sec:mechanism}.
Section~\ref{sec:plasma} presents the data
showing that a plasma is present in the thermal Rydberg gas.
Based on collisional cross-sections
and motional dephasing in an electric field distribution,
we have developed a microscopic model 
which is detailed in Sec.~\ref{sec:model}. 
We confirm our model by comparing measured and simulated data
before concluding the manuscript with Sec.~\ref{sec:conclusion}.
%
\section{Setup and Methods}
\label{sec:experiment}
\begin{figure}[b]
	\includegraphics[]{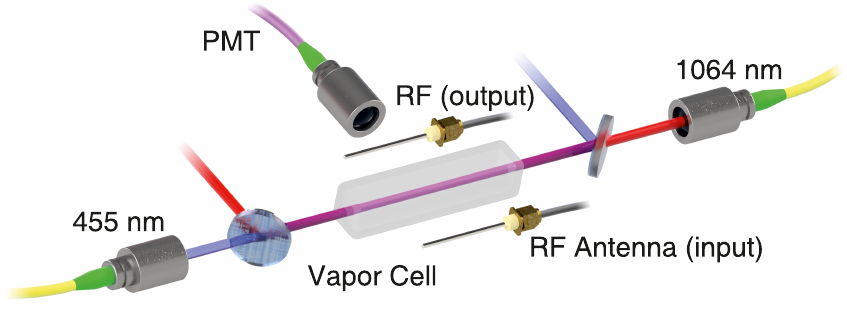}
    \caption{Experimental setup.
    Two counter propagation laser beams excite the Cs atoms in a vapor cell to a Rydberg state.
    The change in the radio-frequency transmission through the atomic sample
    is measured with the dipole antennas shown in the figure. 
	The atomic fluorescence is captured with a fiber, bandpass-filtered and detected on a PMT.}
    \label{fig:setup}
\end{figure}

\begin{figure}[htbp]
    \includegraphics[]{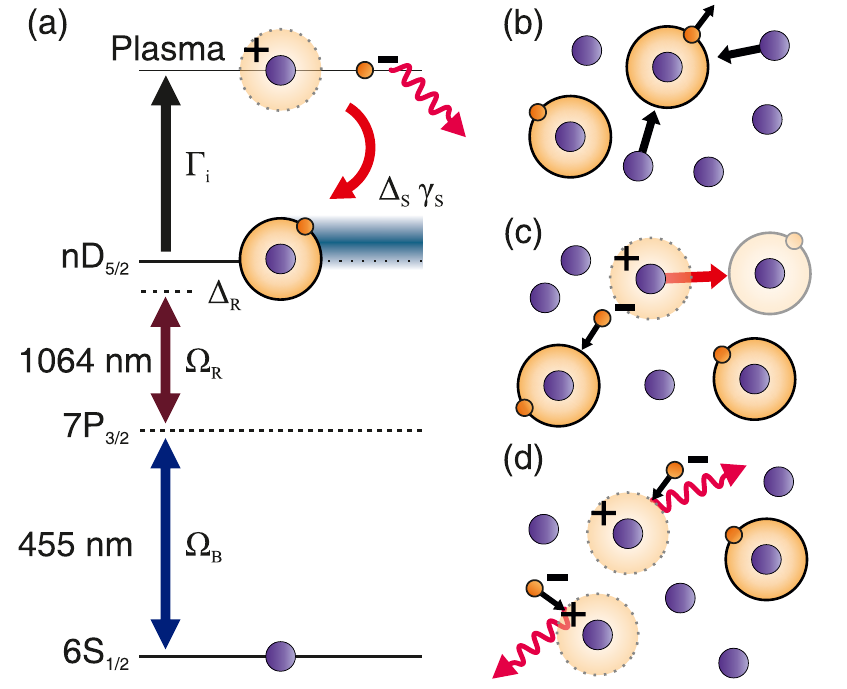}
    \caption{(a) 3-Level diagram for optical excitation
    with ionization to an additional plasma state.
	The plasma causes a shift and broadening of the Rydberg level,
	and recombination results in fluorescence.
	(b)-(c) Illustration of the processes leading to plasma formation and bistability,
	as described in Sec.~\ref{sec:mechanism}.
	(b) Rydberg-ground state collisions ionize some atoms leading to production of fast electrons.
	(c) Avalanche ionization due to fast electrons
	and broadening due to steep electric field gradients
	can facilitate higher Rydberg population.
	(d) Fluorescence due to recombination in the plasma.}
    \label{fig:schematic}
\end{figure}

Figure \ref{fig:setup} shows the experimental setup
for the investigation of the origin of the bistability
and the analysis of the plasma.
A vapor cell containing cesium is placed in the path of two counter propagating,
collimated laser beams.
Each beam has a $1/e^2$ radius of \SI{0.7}{\milli\meter}.
The vapor cell has a square cross-section with a side length of \SI{1}{\centi\meter}.
The length in the direction along the beam is \SI{5}{\centi\meter}.
The bulk of the cesium resides in a reservoir
attached to a side arm of the vapor cell.
The temperature of the reservoir is stabilized
to control the vapor density inside the vapor cell.
The main body of the vapor cell is held at a higher temperature
to prevent alkali condensation on the windows.
We determine the cesium atomic density for each measurement
by recording and fitting the absorption profile of the cesium D$_2$ spectrum.
The densities used for the experiments are varied from
\SI{5e10}{\per\cubic\centi\meter} to \SI{7e12}{\per\cubic\centi\meter}.
We excite Cs atoms to the $n\mathrm{D}_{5/2}$ Rydberg state ($n=30$ and 42)
via the inverted wavelength two-photon scheme depicted in Fig.~\ref{fig:schematic}(a).
The first laser is a frequency-stabilized 455-nm laser
-- a frequency doubled diode laser --
tuned to the cesium 6S$_{1/2}, F = 3 \rightarrow$ 7P$_{3/2}, F' = 4$ transition.
Rabi frequencies, $\Omega_\mathrm{B}/2\pi$, used for the experiments
range between \SI{1}{\mega\hertz} and \SI{12}{\mega\hertz}.
The 455-nm laser is locked via
dichroic atomic vapor spectroscopy \cite{corwin1998frequency}
in a separate cesium vapor cell.
The second laser -- a fiber-amplified 1064-nm diode laser --
is either scanned over the cesium 
7P$_{3/2} \rightarrow$ $n\mathrm{D}_{5/2}$ transition
with a detuning $\Delta_\mathrm{R}$
(measurements with $n = 42$),
or locked on resonance
(measurements with $n = 30$),
using another cesium vapor cell.
Side-of-fringe locking~\citep{fritschel1989frequency}
to the reference electromagnetically induced transparency (EIT)
signal is used for this light field.
Rabi frequencies, $\Omega_\mathrm{R}/2\pi$, are varied
between \SI{3}{\mega\hertz} and \SI{90}{\mega\hertz}.
The laser detuning scans
are calibrated using a Fabry-P\'erot interferometer
with a free-spectral range of $\Delta\nu_\mathrm{FSR} = \SI{1.5}{GHz}$
in combination with an EIT reference signal.
The fluorescence from the highly excited atoms
was used to identify the formation of the plasma.
Due to the high laser powers involved,
measured changes in the transmission of either light field
suffer from a large background signal.
For this reason,
we either measure the change in transmission of a radio-frequency field
(frequency f$_\mathrm{RF}$ ranging from \SIrange{8}{300}{\mega\hertz})
through the vapor cell in the presence of Rydberg excitations and ions, 
or the change in fluorescence to obtain a signal. 
Long- and short-pass filters are placed in front of a photo-multiplier tube (PMT)
with a bandwidth of $\SI{20}{\kilo\hertz}$
so that the fluorescence light between \SIrange{500}{800}{\nano\meter}
transverse to the laser beam propagation direction
can be collected.
With this interval,
we exclude the 455-nm laser light,
and the strongly fluorescing decay via the D2 line at \SI{852}{\nano\meter}
that can already occur with the 455-nm excitation on its own.
The fluorescence is collected by an imaging system
consisting of two lenses with focal lengths of
$f_1 = \SI{30}{\milli\meter}$ and $f_2 = \SI{8}{\milli\meter}$.
The second lens focuses the light into a multi-mode fiber,
\SI{400}{\micro\meter} core diameter.
The fluorescence is imaged on to the fiber facet
from within the excitation laser beam path,
which is about \SI{45}{\milli\meter} in front of the first lens.
The diameter of the spot from which photons are collected
is approximately \SI{1.5}{\milli\meter}.
Overall,
we estimate that the collection efficiency is $\sim 0.1\%$
of the photons emitted into all directions from that spot.
For the transmission measurements
of the radio-frequency fields,
the RF field is coupled into the vapor cell
with a \SI{4}{\centi\meter}-long dipole antenna
that is oriented parallel to the laser beams.
A second,
identical antenna at the opposite side of the vapor cell picks up the signal.
In some measurements,
to appear later in the paper,
we measure the atomic fluorescence
as a surrogate for the microwave transmission measurements.
We show that this proxy is a valid one.
In order to improve the signal-to-noise ratio,
we modulate the radio-frequency amplitude as well as the laser intensity
and make use of lock-in amplification.
We used two lock-in steps
in combination with quadrature detection.
\section{Bistability Mechanism}
\label{sec:mechanism}
In previous work on optical bistability in thermal vapor cells
the authors assumed that mutual Rydberg atom interactions embody the mechanism.
The effect of charges was not incorporated
into the model used to explain the observed behavior
\citep{carr2013nonequilibrium,vsibalic2015driven,de2016intrinsic}
The term \emph{optical bistability},
in the context of Rydberg atoms and plasmas in thermal vapor cells,
refers to the appearance of a hysteresis
in the observed spectral response of the system
when scanning one laser in a multi-photon excitation scheme
across a Rydberg state resonance.
The hysteresis arises due to the competition between a non-linear shift
-- in our case caused by a Stark shift due to surrounding charges --
on the one side
and decay of the excited state population on the other side.

\begin{figure}[htbp]
    \includegraphics[]{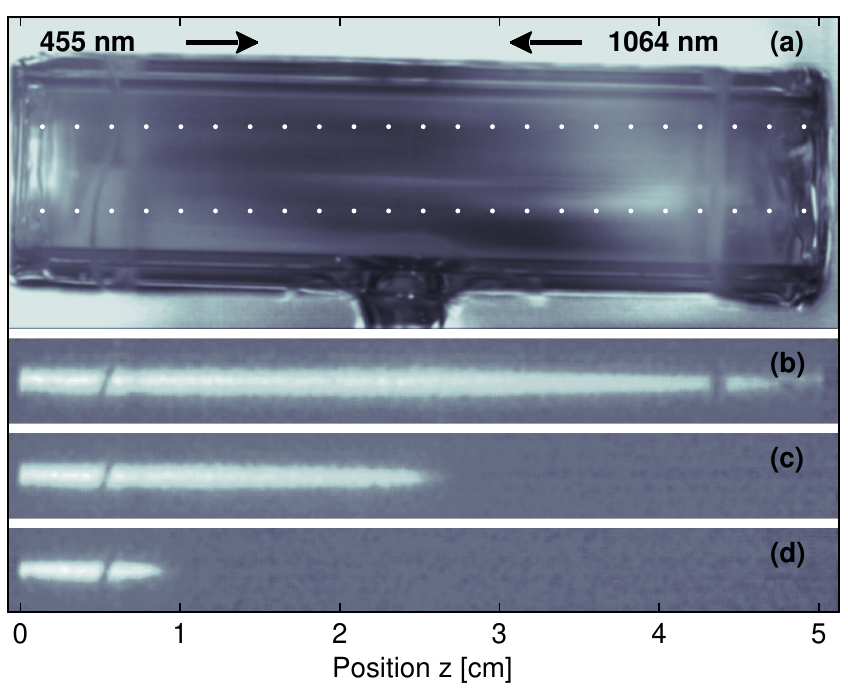}
    \caption{(a) Side-view of the cell. 
    The lasers propagate between the dotted lines, 
    455-nm laser from left to right,
    1064-nm laser from right to left.
    (b-d) Fluorescence image for wavelengths between
    500 and \SI{800}{\nano\meter} with background subtracted.
    Scanning the Rydberg laser across resonance
    creates a \emph{lightsaber-like} fluorescing ray.
    When appearing, 
    the glowing beam extends across the whole cell at once (b), 
    and then gradually retracts from right to left as the detuning is scanned (c, d),
    before smoothly vanishing completely.              
    The 455-nm laser intensity decreases from left to right
    due to absorption.
    At the transition between light and dark side,
    the conditions for plasma formation are not fulfilled any further along the beam.
    The defects on the left and right hand side are due to quartz sublimate on the surface of the cell.
    The position $z$ along the cell is indicated on the bottom axis.
    }
    \label{fig:cell}
\end{figure}

Figure \ref{fig:cell}(a) shows a photograph of the vapor cell from the side.
The beams travel horizontally through the cell. 
Given suitable combination of density, laser power and detuning;
the lasers create a fluorescing ray (Fig.~\ref{fig:cell}(b-d)).
As the 1064-nm laser is scanned back and forth across the Rydberg resonance position,
fluorescence suddenly appears to streak across the vapor cell for one scan direction,
but then gradually retracts towards the entrance window of the blue laser beam
until it completely vanishes for the opposite scan direction.
The directionality of the phenomenon depends on the sign of the polarizability.
This \emph{lightsaber}-like phenomenon \citep{bowden2016s}
with a distinct transition between a light and a dark side along the beam
(Fig.~\ref{fig:cell}(b-d)) repeats for every cycle of the detuning scan,
and is visible by eye.
This is a visual manifestation of the optical bistability.
In this work,
we modify the overall concept for the occurrence of the hysteresis
as depicted in Fig.~\ref{fig:schematic}.
As evidenced in \citep{weller2016charge},
charges play an essential role in the optical bistability
that takes place in dense thermal Rydberg gases.
In fact,
it is not Rydberg-Rydberg interactions that mediate the optical bistability,
but the plasma of ions and electrons that are created
at the Rydberg densities required to observe the effect.
We show that in typical optical bistability experiments
over a range of parameters where the phenomena can be observed,
a weakly coupled plasma exists in the vapor.
The sudden switching to the plasma state via an avalanche ionization process
leads to the observation of the hysteresis that motivates the term optical bistability.
Here, the term bistability is used in that sense that the system jumps from one metastable state to another via this avalanche dynamics
\citep{letscher2017bistability}.
We have identified two mechanisms that contribute to the ionization
initiating the plasma formation.
Rydberg atoms collide with ground state atoms
(Fig.~\ref{fig:schematic}(b)),
creating ions and free electrons.
The electrons then collide with other Rydberg atoms
(Fig.~\ref{fig:schematic}(c)),
giving rise to a much higher ionization
\mbox{rate $\Gamma_\mathrm{i}$}.
This effect is commonly known as the \emph{avalanche process}.
\citep{vitrant1982rydberg,killian1999creation,robinson2000spontaneous}.
The plasma plays a fundamental role in the bistability observed in the thermal Rydberg gas.
With charges present in the gas,
the Rydberg energy level is shifted due to the Stark shift, $\Delta_\mathrm{S}$.
As thermal atoms move through an electric field distribution
\citep{holtsmark1919verbreiterung},
their energy levels rapidly change
during the atom-light interaction
causing a homogeneous broadening.
Therefore,
an additional dephasing $\gamma_\mathrm{S}$ of the Rydberg population
needs to be taken into account which changes the excitation dynamics of the driven system.
Plasma recombination gives rise
to the distinct fluorescence spectrum
\citep{agnew1968identification,carr2013nonequilibrium,wade2018terahertz}
which can be measured.
The recombination events lead to a broad, discrete distribution
of Rydberg principal quantum number states
which then subsequently decay
(Fig.~\ref{fig:schematic}(d)). 

\section{Plasma Characterization}
\label{sec:plasma}
\begin{figure}[htbp]
    \includegraphics[]{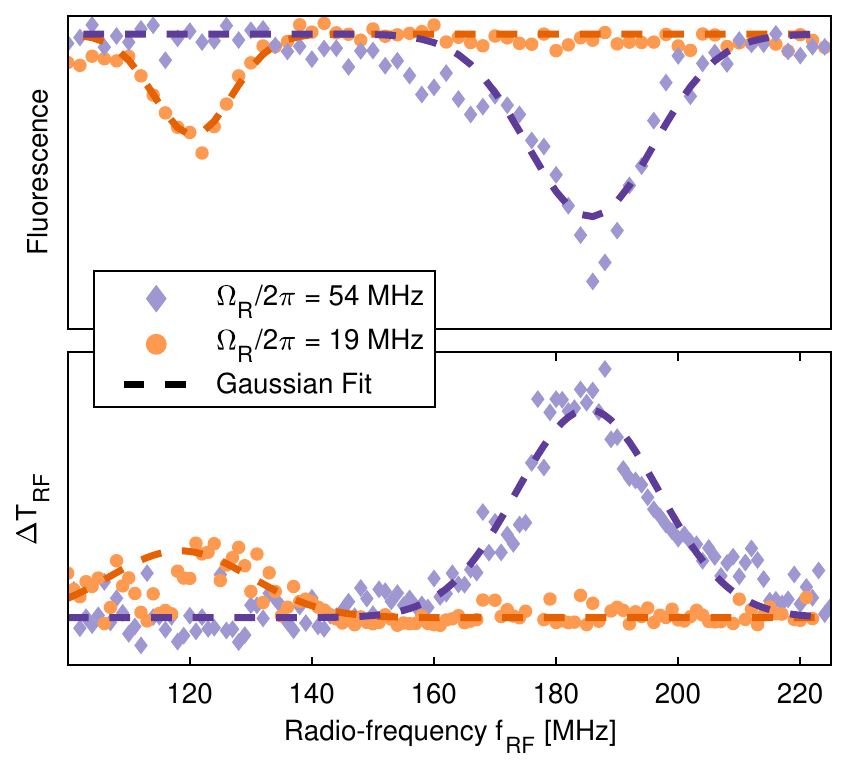}
    \caption{Fluorescence vs.\,radio-frequency (top)
    	and radio-frequency transmission change (bottom)
    	for the 30D$_{5/2}$ Rydberg state
    	for the same experimental conditions, with both lasers locked.
        Experimental parameters are $P_\mathrm{RF}=\SI{-15}{\dBm}$,
		$\Omega_\mathrm{B}/2\pi=\SI{3.6}{\mega\hertz}$
		and $\mathcal{N}_\mathrm{g}= \SI{3e12}{\per\cubic\centi\meter}$.
    	The dashed lines are Gaussian fits.
        }
    \label{fig:transmission:fluorescence}
\end{figure}

\begin{figure}[hbtp]
    \includegraphics[]{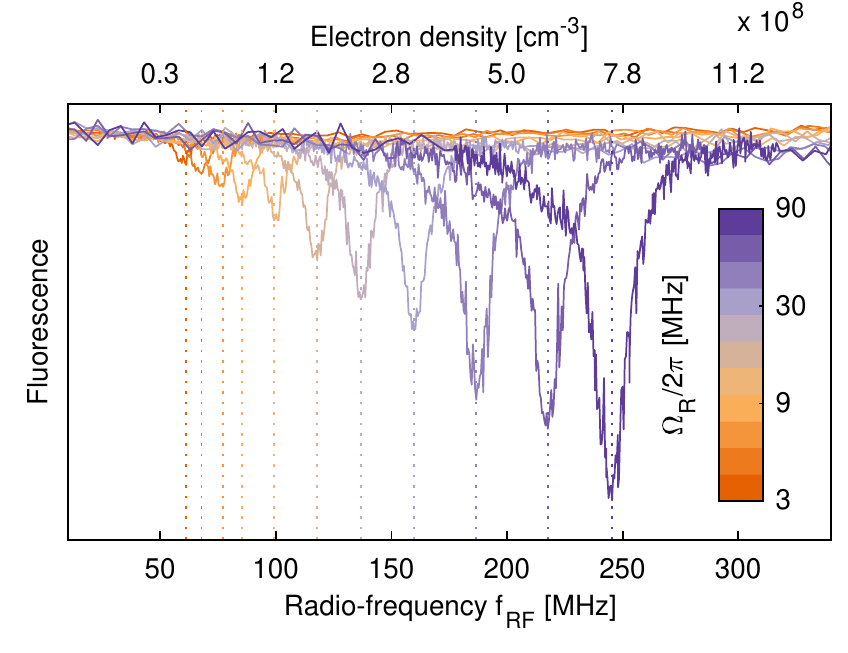}
    \caption{Plasma resonance peaks in the fluorescence signal. 
    Experimental parameter: radio-frequency power $P_\mathrm{RF} = \SI{-16}{\dBm}$,
    density $\mathcal{N}_\mathrm{g} = \SI{3e12}{\per\cubic\centi\meter}$,
    $\Omega_\mathrm{B}/2\pi = \SI{3.6}{\mega\hertz}$.
    The 30D$_{5/2}$ Rydberg state was used for the measurements.
    }
    \label{fig:plasmafrequency}
\end{figure}

\begin{figure}[htbp]
    \includegraphics[]{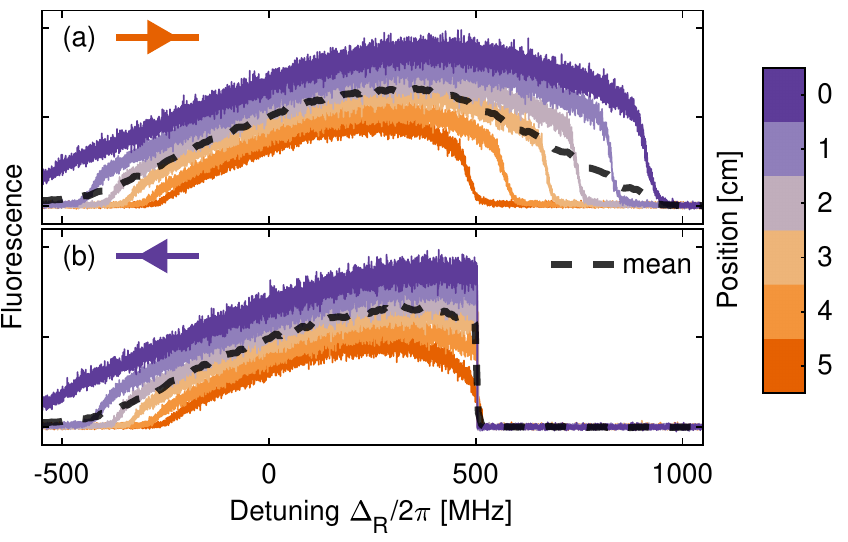}
    \caption{Fluorescence signal collected from different spatial positions along the vapor cell,
    with increasing distance from the entrance window of the blue laser beam. 
    (a) Scan from red to blue detuning (i.e.,~left to right).
    (b) Scan from blue to red detuning.  
    All traces in (b) feature the same sharp edge at approx.~$\SI{500}{\mega\hertz}$.
    The dashed line shows the mean of the fluorescence signal.
    The mean is similar to the integrated signal 
    one would obtain by measuring the probe transmission, 
    e.g.,~as in \cite{weller2016charge,carr2013nonequilibrium}.
    Experimental parameters are
    $\mathcal{N}_\mathrm{g} = \SI{3e12}{\per\cubic\centi\meter}$,
    $\Omega_\mathrm{B}/2\pi = \SI{2.5}{\mega\hertz}$ and
	$\Omega_\mathrm{R}/2\pi = \SI{18}{\mega\hertz}$.
	Data is shown for the 42D$_{5/2}$ Rydberg state.
    }
    \label{fig:positions}
\end{figure}

We measured the change in transmission
$\Delta \mathrm{T}_\mathrm{RF}$
of a radio-frequency signal
to show that a plasma exists in the gas.
The results are shown in Fig.~\ref{fig:transmission:fluorescence}.
The resonance position and its shift with the Rydberg density
already suggests the presence of an (electron) plasma.
The effect of the radio-frequency excitation
would not occur in a neutral cesium gas
at these frequencies.
The data-set was taken
for two different Rydberg excitation laser Rabi frequencies, $\Omega_\mathrm{R}$,
but otherwise identical configurations.
Both lasers were locked on resonance
and excite the atoms to the 30D$_{5/2}$ state.
The ground state cesium density was
$\mathcal{N}_\mathrm{g} = 3 \times 10^{12}\,$cm$^{-3}$
and the radio-frequency power delivered to the antenna was 
$P_\mathrm{RF}=\SI{-15}{\dBm}$.
A careful look at Fig.~\ref{fig:transmission:fluorescence}
appears puzzling at first glance
since the radio-frequency field doesn't appear in absorption
but increased transmission.
The enhanced coupling between the two antennas
is justified by the increased electric susceptibility,
$\epsilon > 1$,
and the fact that the antennas are in the near-field of each other.
Effectively, the distance between the antennas is being reduced
by the free charges within the plasma.
Figure~\ref{fig:transmission:fluorescence} additionally shows
the change in fluorescence measured as a function of radio-frequency, $f_\mathrm{RF}$.
The overlap of the radio-frequency and fluorescence signals show
that the fluorescence can be used
as a surrogate for the radio-frequency transmission signal.
Such a connection is important
because the fluorescence signal is much larger
than the radio-frequency transmission signal
and therefore easier and much faster to measure.
The change in fluorescence can be explained
by considering how the radio-frequency field accelerates the free electrons
into an oscillating motion with an amplitude
depending on the frequency of the radio-frequency field.
If the radio-frequency field is resonant with the plasma frequency,
the amplitude of the electron oscillation is maximal
and the motion effectively decreases the electron density within the laser beams.
If the amplitude of the electrons exceeds the radial extent of the laser beams,
the recombination probability in the center of the beam,
where the ions are presumably located and most fluorescence is collected,
decreases.
The electrons can also collide with the walls of the vapor cell,
further reducing the number of particles contributing to the fluorescence signal.
Figure~\ref{fig:plasmafrequency} shows a typical series of measurements
with varying Rydberg excitation laser intensity.
As the Rabi frequency $\Omega_\mathrm{R}$ gradually increases,
the plasma resonance clearly shifts its position and increases in amplitude.
The shift in the plasma resonance enables us to extract the density of the electron plasma.
Given the relation between plasma frequency $f_\mathrm{p}$
and the electron density $\mathcal{N}_\mathrm{e}$,
\begin{equation}
\mathcal{N}_\mathrm{e} = \left(2\pi f_\mathrm{p}\right)^2\frac{\epsilon_0m_\mathrm{e}}{\mathrm{e}^2} ,
\end{equation}
the electron density can be calculated
using the electron mass $m_\mathrm{e}$,
the electronic charge $e$,
and the permittivity of free space $\epsilon_0$.
The axis ticks above Fig.~\ref{fig:plasmafrequency}
are computed with this formula,
yielding an electron density on the order of \SI{1e8}{\per\cubic\centi\meter}.
Note that the observed plasma resonance in Fig.~\ref{fig:plasmafrequency} must be that of the electrons
because a calculation of the ion density using these resonant frequencies
gives densities that are 2 to 100 times larger than the peak cesium vapor density
$\mathcal{N}_\mathrm{g} = \SI{3e12}{\per\cubic\centi\meter}$ we measured in the vapor cell.
Along with the plasma frequency, 
the Debye length,
\begin{equation}
	\lambda_\mathrm{D} =
    \sqrt{\frac{\epsilon_0 k_\mathrm{B} T}{\mathcal{N}_\mathrm{e} e^2}},
\end{equation}
and the Coulomb coupling parameter,
\begin{equation}
	\Gamma_e =
	\frac{\mathrm{e}^2}{4\pi\epsilon_0 k_\mathrm{B} T}
    \sqrt[3]{\frac{4\pi \mathcal{N}_\mathrm{e}}{3}},
\label{eq:couplingparameter}
\end{equation}
fully describe the plasma.
Assuming a plasma temperature equal to the vapor temperature $T = \SI{370}{\kelvin}$,
the Debye length is $\lambda_\mathrm{D} \sim \SI{100}{\micro\meter}$ and
the Coulomb coupling parameter is $\Gamma_e \sim 0.006$. $k_\mathrm{B}$ is the Boltzmann constant. The plasma parameters
indicate that the electron plasma is in the weakly coupled regime.
The associated plasma parameter
\begin{equation}
    N_\mathrm{D} = 
	    \mathcal{N}_\mathrm{e} \frac43\pi\lambda_\mathrm{D}^3,
\end{equation}
which is the number of electrons in a Debye sphere, is $N_\mathrm{D} = 10^3$.
The volume of the Debye sphere contains \SI{1.3e7}{} Cesium atoms.  
The plasma is weakly ionized
and has properties comparable to the earth's ionosphere
\cite{schunk2009ionospheres}.
Figure~\ref{fig:positions} shows a set of fluorescence signal data
for both red to blue and blue to red 1064-nm laser scans at
different fluorescence collection positions $z$ along the vapor cell.
The fluorescence is associated with population in a large number of Rydberg states
because of recombination.
The fluorescence was dispersed with a spectrometer
to verify that a broad distribution of Rydberg states were present.
As evident from the plots,
the system stays in the state with a high Rydberg density
\footnote{High Rydberg density implies high charge density}
when scanning from red towards blue detuned wavelengths, 
even for larger detunings, Fig.~\ref{fig:positions}(a).
The behavior of the curves in Fig.~\ref{fig:positions}(a)
is to be contrasted with that in Fig.~\ref{fig:positions}(b)
where a sharp transition occurs at the same 1064-nm laser detuning
for all fluorescence collection points.
It is apparent that for larger distances from the 455-nm laser beam input window,
corresponding to larger propagation distances of the blue light through the vapor cell,
the overall strength of the fluorescence signal decreases.
The decreasing 455-nm laser intensity affects the detuning 
at which the falling edge of the hysteresis occurs for red to blue detuning,
Fig.~\ref{fig:positions}(a).
In contrast,
the plots in Fig.~\ref{fig:positions}(b) all feature the same sharp edge
at approximately $\SI{500}{\mega\hertz}$.
The rising edge,
Fig.~\ref{fig:positions}(b),
where the data is acquired for blue to red detuning,
is unaffected by the varying 455-nm laser absorption
along the length of the vapor cell.
When the system is in the bistable part of the spectrum,
but in the low Rydberg population state,
the conditions for the ionization only need to be fulfilled
at a single position along the laser beam.
The avalanche ionization then quickly spreads across the whole cell.
We estimate the timescale to be on the order of \SI{10}{\milli\second}.
Typically,
the conditions for the avalanche ionization are fulfilled first
where the maximum Rabi frequency $\Omega_\mathrm{B}$ exists,
i.e.,~at the entrance window of the blue laser.
The prominent jump in the Rydberg population and associated change in fluorescence
is referred to as the \emph{edge frequency}.
In the numerical evaluation of the data,
we determine the edge frequency as the 1064-nm laser detuning value,
$\Delta_\mathrm{E}$,
where the fluorescence signal exceeds $1/4$ of the maximum value in each trace,
when tracing the data from the blue side.
It is useful to look at $\Delta_\mathrm{E}$ to determine
how the laser Rabi frequencies and density scale
with its value in order to learn about the driven system dynamics.
The plasma and its role in bistability can be further analyzed
by investigating the dependence on the laser Rabi frequencies and vapor density.
To study the optical bistability 
the 1064-nm laser is scanned across the 
$7\mathrm{P}_{3/2} \rightarrow 42\mathrm{D}_{5/2}$ transition.
We collected the fluorescent light as described in Sec.~\ref{sec:experiment}.
Since the system is bistable,
both positive and negative scan directions are recorded and treated separately.
We measured spectra for a wide range of experimental parameters,
varying vapor density and both excitation laser Rabi frequencies,
$\Omega_\mathrm{B}$ and $\Omega_\mathrm{R}$.
We focused on analyzing the transition to the plasma state
as we sweep the 1064-nm laser detuning across the spectrum.
We show that it is possible to combine the experimental parameters
into 2 scaling parameters $S$ or $S^\star$,
for the blue to red scan and red to blue, respectively.
The scaling parameter $S$ and $S^\star$ are defined as
\begin{equation}
	S^{\left(\star\right)} = 
    \left(
    	\frac{\mathcal{N}_\mathrm{g}}{\mathcal{N}_\mathrm{0}}
    \right)^a
    \cdot
    \left(
    	\frac{\Omega_\mathrm{B}^{\left(\star\right)}}{\Omega_\mathrm{0}}
     \right)^b
     \cdot 
     \left(
     	\frac{\Omega_\mathrm{R}}{\Omega_\mathrm{0}}
     \right)^c  .
    \label{eq:scaling}
\end{equation}
The two sets of exponents $\left(a, b, c\right)$
are chosen in such a way
that the frequencies where the system jumps
between high and low Rydberg excitation 
linearly scale with $S^\star$ or $S$.
The asterisk denotes that the Rabi frequency
is \emph{adjusted} for absorption up to the position
where the fluorescence measurement is made,
see Appendix~\ref{apx:absorption}.
Previous experiments demonstrate
that the frequency shift of the phase transition
additionally scales with the forth power
of the effective principal quantum number,
$n^{\star 4}$~\citep{de2016intrinsic}.
This is consistent with our interpretation,
because the geometric cross-section of Rydberg atoms 
for ionizing collisions also scales with $n^{\star 4}$.

\begin{figure}[htbp]
    \includegraphics[]{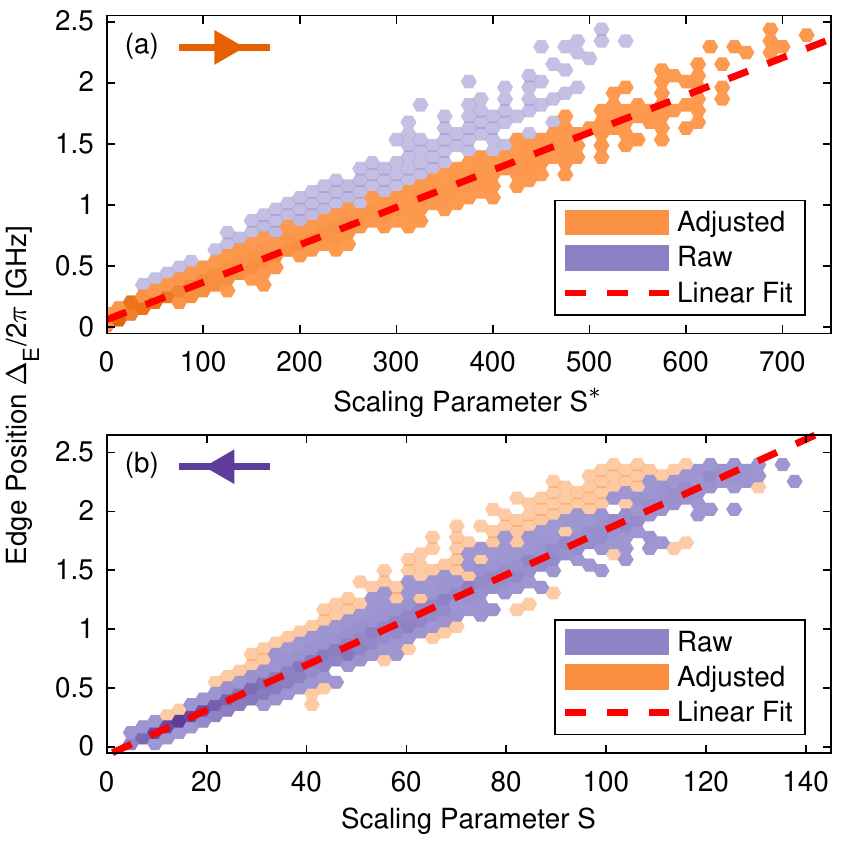}
    \caption{
    Map of the bistability frequency edge,
    $\Delta_\mathrm{E}$ as a function of the scaling parameters $S$ and
    $S^{\star}$, 
    according to Eq.~\ref{eq:scaling}.
    (a) Red to blue detuning.
    (b) Blue to red detuning.
    For comparison,
    the larger spread of the distributions with the respective 
    raw and absorption adjusted Rabi frequencies are shown.}
    \label{fig:universal}
\end{figure}

The set of data points we acquired for the bistability edge position
are plotted in Fig.~\ref{fig:universal}.
The plots cover $\Omega_\mathrm{B}/2\pi$
over a range of \SIrange{1}{12}{\mega\hertz},
$\Omega_\mathrm{R}/2\pi$ from \SIrange{15}{50}{\mega\hertz},
and $\mathcal{N}_\mathrm{g}$ within
\SIrange{5e10}{7e12}{\per\cubic\centi\meter}.
The values for the three scaling exponents $\left(a, b, c\right)$ are obtained 
by a non-linear least-squares fit to the data plotted in Fig.~\ref{fig:universal},
for each scan direction.
The results are given in Tab.~\ref{tab:scaling}.
The fact that the values are different for $S$ and $S^\star$
indicates that the dynamics are different for the two detuning cases
as is expected for bistable behavior.
\begin{table}[htbp]
\setlength{\tabcolsep}{6pt} 
\renewcommand{\arraystretch}{1.25} 
\begin{tabular}{lc|c|c|c}
  & (Scan) & $a$ ($\mathcal{N}_\mathrm{g}$) & $b$ ($\Omega_\mathrm{B}$)& $c$ ($\Omega_\mathrm{R}$) \\
    \hline
		$S$ & $\left(\leftarrow\right)$ & $0.81 \pm 0.01$ & $0.95 \pm 0.02$ & $1.08 \pm 0.02$\\
		$S^\star$ & $\left(\rightarrow\right)$ & $0.54 \pm 0.01$ & $0.56 \pm 0.01$ & $0.97 \pm 0.02$ 
    \end{tabular}
    \caption{Scaling exponents as in Eq.~\ref{eq:scaling} with 95\% confidence intervals.}
    \label{tab:scaling}
\end{table}
Interpreting the numerical values of the exponents for the scaling parameters
reveals interesting properties of the bistable system.
An intuitive picture
based on rate equations
can be developed,
to explain the rough trend of the determined exponents,
$(a,b,c)$.
The most important parameters are the density of Rydberg atoms and ions,
$\mathcal{N}_\mathrm{Ryd}$ and $\mathcal{N}_\mathrm{ion}$.
The two steady state densities
can be obtained by solving the coupled system of rate equations,
\begin{equation}
\begin{aligned}
  \dot{\mathcal{N}}_\mathrm{ion} &=&
    &\mathcal{N}_\mathrm{Ryd} \mathcal{N}_\mathrm{g} \sigma_\mathrm{g}  \overline{v}
    + \mathcal{N}_\mathrm{Ryd} \mathcal{N}_\mathrm{ion} \sigma_\mathrm{e} \overline{v}' \\
    & & -&\mathcal{N}_\mathrm{ion}\mathcal{N}_\mathrm{e} \sigma_\mathrm{r} \overline{v}''
    -\mathcal{N}_\mathrm{ion} \Gamma_\mathrm{t}\\
  \dot{\mathcal{N}}_\mathrm{Ryd} &=&
    -&\mathcal{N}_\mathrm{Ryd} \mathcal{N}_\mathrm{g} \sigma_\mathrm{g}  \overline{v}
    -\mathcal{N}_\mathrm{Ryd} \mathcal{N}_\mathrm{ion} \sigma_\mathrm{e} \overline{v}' \\
    & & +&\mathcal{R}_\mathrm{pump} .
\end{aligned}
\label{eq:rate}
\end{equation}
The cross-sections for
Rydberg-ground state collisions $\sigma_\mathrm{g}$,
Rydberg-electron collisions $\sigma_\mathrm{e}$,
and recombination of ions with electrons $\sigma_\mathrm{r}$ and
mean relative velocity $\overline{v}$ of the respective 
interacting species are parameters in these equations.
Ions diffuse away at a rate $\Gamma_\mathrm{t}$.
The effective rate at which atoms are excited to the Rydberg state
is $\mathcal{R}_\mathrm{pump}$
The decay of the Rydberg population
results in an offset in the resulting equations,
and does not play a role for the scaling behavior
in the parameters of interest.
Depending on whether the system approaches $\Delta_\mathrm{E}$
from the ground state side
($\mathcal{N}_\mathrm{ion}, \mathcal{N}_\mathrm{e}, \mathcal{N}_\mathrm{Ryd} \approx \SI{0}{\per\centi\meter\cubed}$)
or the highly populated excited state side
($\mathcal{N}_\mathrm{ion}, \mathcal{N}_\mathrm{e}, \mathcal{N}_\mathrm{Ryd} \gg \SI{0}{\per\centi\meter\cubed}$),
different approximations for Eq.~\ref{eq:rate} can be applied.
For the analysis of the scaling behavior in both of these two extremes
in the following paragraphs,
we assume the electron and ion density to be approximately the same,
$\mathcal{N}_\mathrm{e} \approx \mathcal{N}_\mathrm{ion}$,
both given by the parameter $\mathcal{N}_\mathrm{c}$.

For blue to red detuning sweeps which are described by
$S$ in Tab.~\ref{tab:scaling} and Fig.~\ref{fig:universal}(b),
$\Delta_\mathrm{E}$ is approximately linear with respect to all 3 parameters,
$\left(a, b, c\right) \approx 1$.
In order for the system to jump to the plasma state,
it is sufficient to trigger the threshold for the avalanche mechanism at a single point along the vapor cell.
All traces in Fig.~\ref{fig:positions}(b)
feature a sharp edge at the same detuning value.
It is important to understand
that the Stark shift and associated spectral broadening due to the charge distribution
does \emph{not} play a role in this case. 

The system starts from the ground state,
so we can approximate the rate equations, Eq.~\ref{eq:rate}, as
\begin{equation}
\begin{aligned}
    \dot{\mathcal{N}}_\mathrm{c} &\stackrel{\mathcal{N}_\mathrm{c} \ll \mathcal{N}_\mathrm{g} }{=}
    \mathcal{N}_\mathrm{Ryd} \mathcal{N}_\mathrm{g} \sigma_\mathrm{g}  \overline{v}''
    -\mathcal{N}_\mathrm{c} \Gamma_\mathrm{t} \\
    \dot{\mathcal{N}}_\mathrm{Ryd} &\stackrel{\mathcal{N}_\mathrm{c} \ll \mathcal{N}_\mathrm{g}}{=}
    - \mathcal{N}_\mathrm{Ryd} \mathcal{N}_\mathrm{g}
    \sigma_\mathrm{g} \overline{v}
    + \mathcal{N}_\mathrm{g}
    \frac{\Omega_\mathrm{B}\Omega_\mathrm{R}}{
    \stackrel{\sim}{\Delta}} .
\end{aligned}
\label{eq:nll1} 
\end{equation}
Here, the pump rate to the Rydberg state is taken to be
the effective two-photon Rabi frequency divided by the effective detuning
$\stackrel{\sim}{\Delta}$, making this a non-resonant 2-photon coherent excitation.
Without the plasma or a significant number of charges,
the line width of the Rydberg transition corresponds to its natural width,
and both lasers are above saturation.
Hence,
the Rabi frequencies contribute linearly to the pump rate.
In steady state,
the charge density from Eq.~\ref{eq:nll1} is given by
\begin{equation}
    \mathcal{N}_\mathrm{c} \propto
    \mathcal{N}_\mathrm{g} \Omega_\mathrm{B} \Omega_\mathrm{R} ,
\end{equation}
which reflects the measured linear scaling in all three parameters.

For the opposite scan direction,
scaling parameter $S^\star$ in Tab.~\ref{tab:scaling} and Fig.~\ref{fig:universal}(a),
the dominant terms in Eq.~\ref{eq:rate} are
\begin{equation}
\begin{aligned}
    \dot{\mathcal{N}}_\mathrm{c} &\stackrel{\mathcal{N}_\mathrm{c} \gg 0}{=}
    \mathcal{N}_\mathrm{Ryd} \mathcal{N}_\mathrm{c} \sigma_\mathrm{e} \overline{v}
    - \mathcal{N}_\mathrm{c}^2 \sigma_\mathrm{r} \overline{v} \\
    \dot{\mathcal{N}}_\mathrm{Ryd} &\stackrel{\mathcal{N}_\mathrm{c}\gg 0}{=}
    - \mathcal{N}_\mathrm{Ryd} \mathcal{N}_\mathrm{c} \sigma_\mathrm{e} \overline{v}
    + \mathcal{N}_\mathrm{g}
    \frac{\Omega_\mathrm{B}}{2\Gamma_\mathrm{D}}
    \frac{\Omega_\mathrm{R}^2}{\Gamma_\mathrm{Ryd}}.
\end{aligned}
\label{eq:ngg1} 
\end{equation}
With the plasma present,
the Rydberg transition is massively broadened by the varying Stark shifts
and the 1064-nm laser intensity is effectively
\emph{below} the saturation intensity.
As a result,
the Rydberg population depends quadratically on $\Omega_\mathrm{R}$.
The 455-nm laser transition remains saturated.
The intermediate population is proportional to the Rabi frequency $\Omega_\mathrm{B}$
divided by the Doppler width $\Gamma_\mathrm{D}$,
reflecting the fact that more velocity classes
contribute to the Rydberg population because of the
increasing power broadening.
Solving Eq.~\ref{eq:ngg1} for the steady state charge density gives
\begin{equation}
    \mathcal{N}_\mathrm{c} \propto   
    \sqrt{\mathcal{N}_\mathrm{g}}
    \sqrt{\Omega_\mathrm{B}}
    \Omega_\mathrm{R},
\end{equation}
which is in agreement with the experimentally obtained scaling $S^\star$
shown in Tab.~\ref{tab:scaling}.

It is important to note
that most of the measurements were not performed
in one of the extreme cases mentioned here.
Therefore a detailed numerical model
is developed in the next section.
%

\section{Microscopic Model}
\label{sec:model}
\begin{figure*}[t]
    \includegraphics[width=\textwidth]{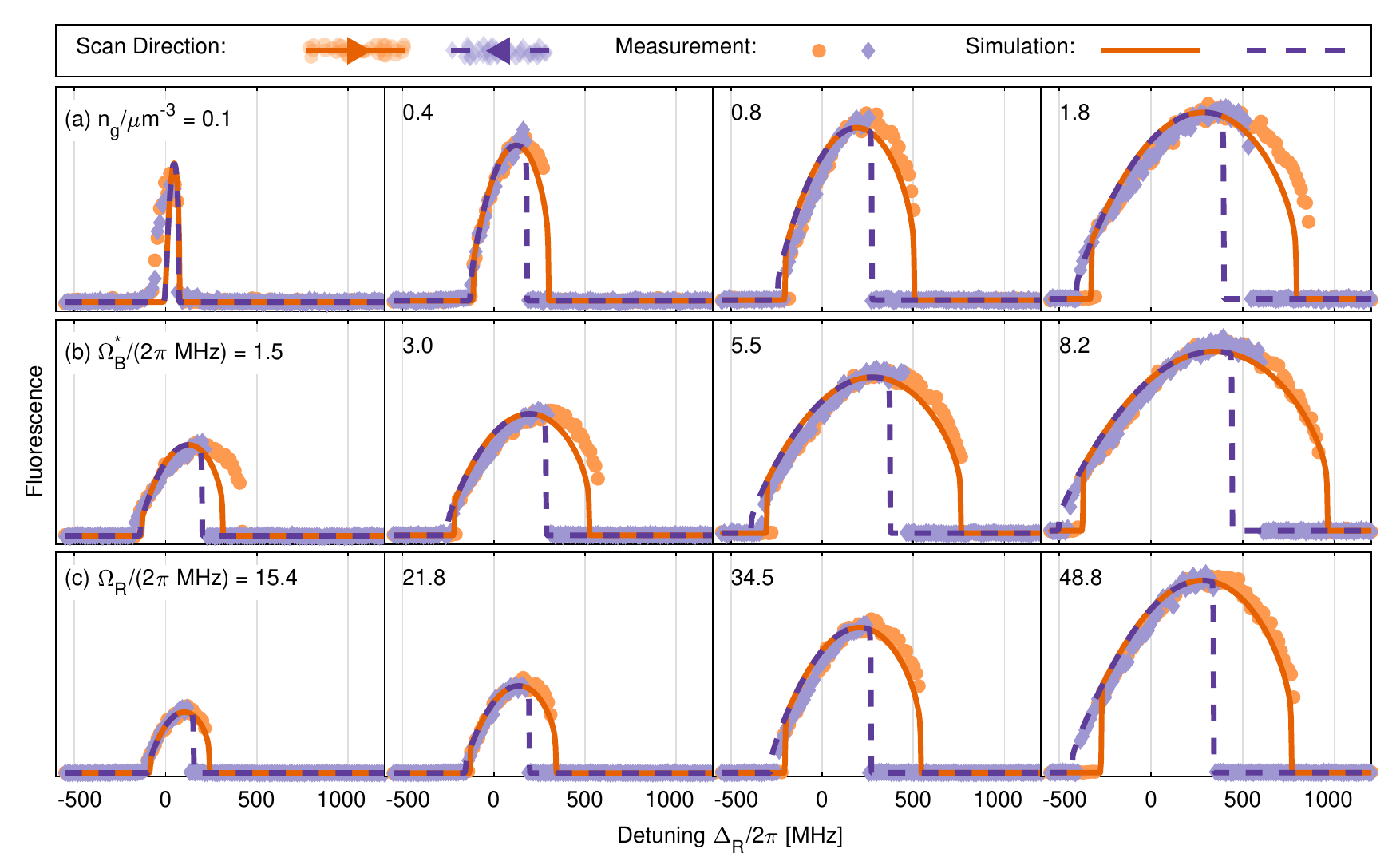}
    \caption{Comparison between fluorescence obtained as a function of 1064-nm laser detuning and calculations. 
    Each row shows the variation of only one experimental parameter: ground state atomic density (a), $\Omega_\mathrm{B}$ (b), and $\Omega_\mathrm{R}$ (c).
    The measurement is shown in orange dots
	(scan from red towards blue wavelengths)
	and purple diamonds (scan from blue towards red wavelengths).
    For the simulation we use orange solid and dashed purple lines,
	respectively.
    Experimental settings are used as input parameter for the simulation.
	Only the fluorescence signal amplitude is adjusted to the data.
	Experimental settings are:
	(a)
	$\Omega_\mathrm{B}/2\pi = \SI{2.6}{\mega\hertz}$, 
	$\Omega_\mathrm{R}/2\pi = \SI{15.4}{\mega\hertz}$
	(b)
	$\Omega_\mathrm{R}/2\pi = \SI{15.4}{\mega\hertz}$,
	$\mathcal{N}_\mathrm{g} = \SI{1.3e12}{\per\cubic\centi\meter}$
	(c)
	$\Omega_\mathrm{B}/2\pi = \SI{5.3}{\mega\hertz}$,
	\mbox{$\mathcal{N}_\mathrm{g} = \SI{0.2e12}{\per\cubic\centi\meter}$}.	
	}
    \label{fig:simulation:measurement}
\end{figure*}

%
To calculate the state populations and coherences
of an ensemble of thermal atoms excited by laser beams,
we find the steady state solution to the Lindblad master equation 
\begin{equation}
	\dot{\rho} = -\frac{i}{\hbar}\left[\rho, H\right] + L, 
    \label{eq:masterequation}
\end{equation}
for all possible velocities,
for a given set of experimental parameters;
i.e., a given set of density, laser detuning and Rabi frequencies.
The Rydberg state population, for example,
is obtained
by averaging over the ensemble
weighted by the Boltzmann probability distribution describing the atomic velocities,
and the electric field distribution
which causes an additional detuning of the Rydberg state due to the Stark shift.
Analogous to Doppler averaging, 
we integrate over the combined probabilities of the electric field and velocity
to obtain the observable quantities from the density matrix.
Our model of the atomic system is depicted in Fig.~\ref{fig:schematic}(a-d).
Applying the density matrix formalism,
we describe the thermal gas interacting with the lasers
as a conventional 3 + 1 level system.
The lower 3 levels  describe the neutral gas
including its excitation to the Rydberg state.
The additional level (+1) describes the generation of ions
(Fig.~\ref{fig:schematic}(a)) which gives rise to the plasma.
Details of the density matrix, Hamiltonian and the Lindblad operator,
can be found in the Appendix~\ref{apx:master}.
We have identified two mechanisms responsible for the ionization of the Rydberg atoms.
The ionization rate,
which depends on the ion population $\rho_\mathrm{ion}$,
is given by
\begin{equation}
	\Gamma_\mathrm{i}\left(\rho_\mathrm{ion}\right) = 
	\mathcal{N}_\mathrm{g} 
	\sqrt{\frac{8 k_\mathrm{B}T}{\pi}}
	\left(
    \sigma_\mathrm{g} 
    \sqrt{\frac{2}{m_\mathrm{Cs}}}
	+ \sigma_\mathrm{e} \frac{\rho_\mathrm{ion}}{\sqrt{m_\mathrm{e}}}
    \right) ,
\end{equation}
where $m_\mathrm{Cs}$ is the cesium mass.
$\sigma_\mathrm{g}$ and $\sigma_\mathrm{e}$
are the ionization cross-sections
for Rydberg atoms colliding with ground-state atoms 
(Fig.~\ref{fig:schematic}(b))
or electrons (Fig.~\ref{fig:schematic}(c)),
respectively.
Since the ground-state density is not significantly influenced by the excitations,
we take $\rho_\mathrm{g} \approx 1$ in our model.
We assume values for $\sigma_\mathrm{g}$ to be on the order of
$0.06 \sigma_\mathrm{geo}$ \citep{vitrant1982rydberg},
with $\sigma_\mathrm{geo}=\pi \left(a_0 n^{\star 2}\right)^2$ being the geometric cross-section
of the Rydberg state with effective principal quantum number $n^\star$.
For $\sigma_\mathrm{e}$ -- to our knowledge -- there is no literature reference for cesium.
However,
according to experiments with sodium Rydberg states \citep{nagesha2003electron}
a reasonable range for this parameter would be $1-10\,\sigma_\mathrm{geo}$.
We neglect that the electron impact ionization cross-section depends
on the velocity of the electron~\citep{vrinceanu2005electron}.
In a similar fashion, the rate
\begin{equation}
	\Gamma_\mathrm{d}\left(\rho_\mathrm{ion}\right) = 
	\Gamma_\mathrm{t} +
	\mathcal{N}_\mathrm{g}
	\sqrt{\frac{8 k_\mathrm{B}T}{\pi m_\mathrm{e}}}\sigma_\mathrm{r}\rho_\mathrm{ion}
	\label{eq:deionization}
\end{equation}
quantifies the mechanisms leading to losses in the charged particle population.
The most obvious contribution
is charges simply leaving the interaction volume due to their thermal motion
which is described by $\Gamma_\mathrm{t}$.
The second term describes recombination of ions with electrons into neutral particles.
The transit time losses, $\Gamma_\mathrm{t}$,
are determined by the diameter of the excitation volume
and the velocity of the particles.
The measured fluorescence rate can be used to estimate the actual recombination rate,
and therefore to determine the recombination cross-section $\sigma_\mathrm{r}$.
The expected value for $\sigma_\mathrm{r}$
can be estimated in semiclassical treatment with Kramer's formula \citep{kramers1923xciii},
which gives a value of $\SI{1e-5}{}\sigma_\mathrm{geo}$.
However,
for the densities shown in the dataset in Fig.~\ref{fig:simulation:measurement},
the approximations in Eq.~\ref{eq:nll1} are not completely valid,
and therefore the transit time losses of the ions,
$\Gamma_\mathrm{t}$ still play a non-negligible role.
For even higher charge densities,
the neighboring 42D$_{3/2}$
is shifted more than the spacing of $2\pi \times \SI{1}{\giga\hertz}$,
and starts to overlap with the spectrum of the 42D$_{5/2}$ state.
Hence, we focused on the parameter space
where the two states are still separated.
As the distinction between a linear and quadratic component
-- such as in in Eq.~\ref{eq:nll1} --
is not feasible in the vicinity of the origin of the plot,
c.f. Fig.~\ref{fig:universal},
we reduce the complexity of the model by the following substitution.
We replace the recombination term in Eq.~\ref{eq:deionization} with
\begin{equation}
	\mathcal{N}_\mathrm{g}
	\sqrt{\frac{8 k_\mathrm{B}T}{\pi m_\mathrm{e}}}\sigma_\mathrm{r}\rho_\mathrm{ion}
	= \Gamma_\mathrm{r}
	= const. ,
\end{equation}
and thereby assume $\Gamma_\mathrm{d}$ to be an effective ion-loss rate,
with a constant value independent of the ion density.
We also suppose ions to leave the beam geometry
at the same rate $\Gamma_\mathrm{t}$ as Rydberg atoms.
Our estimation of the respective rates can be found in Appendix \ref{apx:photonflux}.
The electric field due to the present charges rapidly varies
as the atoms move through the plasma.
A distribution of detunings needs to be considered
because each atom sees a different sequence of local electric fields,
resulting in different accumulated phase shifts.
The random distribution of phase shifts can be treated
as an effective shift and homogeneous broadening due to dephasing of the Rydberg transition.
In our model,
we define the overall dephasing factor as
\begin{equation}
	\gamma = \gamma_0 + \gamma_\mathrm{S}\quad.
\end{equation}
$\gamma_0/2\pi = \SI{1}{\mega\hertz}$
accounts for the laser line width.
The effect of the Stark shift distribution is approximated to be proportional to its weighted mean
\begin{equation}
    \gamma_\mathrm{S} = \overline{\Delta_\mathrm{S}} = 
    \int\limits_{0}^{\infty}
    \mathrm{d}E\quad
    \mathcal{P}_\mathcal{N}\left(E\right)
	\Delta_\mathrm{S}\left(E\right).
	\label{eq:dephasing}
\end{equation}
The details of the electric field distribution are found in the Appendix~\ref{apx:efield}.
The fluorescence spectrum is proportional to the square of the charge density,
$\rho_\mathrm{ion}^2$,
because of two-body recombination of ions and electrons,
and the before mentioned assumption of
$\mathcal{N}_\mathrm{e} \approx \mathcal{N}_\mathrm{ion}$.
This process leads to the broad, discrete distribution
of Rydberg states which then subsequently emit the fluorescence via their decay.
In order to obtain a spectrum to compare to our data,
the solution of Eq.~\ref{eq:masterequation} for the ion state population
of the ensemble of atoms in the vapor cell needs to be calculated.
The analytic expressions for the velocity distribution and Stark shift distribution has to be used
to average the density matrix elements.
The value for a density matrix element for a single laser detuning and charge density is the average of the density matrix element obtained using
the respective probability distribution functions for the Stark shifts and the atomic velocities,
\begin{equation}
    \overline{\rho}_{i,j} = 
    \int\limits_{-\infty}^{\infty}
    \mathrm{d}v
    \int\limits_{0}^{\infty}
    \mathrm{d}E \quad
    f\left(v)\right)
    \mathcal{P}_\mathcal{N}\left(E\right)
    \rho_{i,j}\left(v,E\right).
\end{equation}
$f\left(v\right)$ \citep{maxwell1860v}
is the one dimensional Maxwellian distribution function
for the velocity, $v$, in the beam direction
and $P_\mathcal{N}(E)$ is the electric field distribution as given in Eq.~\ref{eq:holtsmark}.
In order to obtain an equilibrium solution for the charge density,
we make use of an iterative method as described in the Appendix~
\ref{apx:iterative}.
We fit the theory to the spectra found in Fig.~\ref{fig:simulation:measurement}.
The parameters that were fit are  shown in Tab.~\ref{tab:params}.
We have refined their values via non-linear least-squares optimization
using the Levenberg-Marquardt method 
\citep{levenberg1944method,marquardt1963algorithm}.
\begin{table}[htbp]
\setlength{\tabcolsep}{8pt} 
\renewcommand{\arraystretch}{1.5} 
\begin{tabular}{c | c | c | l}
    Var. & Fit result & Expect. coeff. & Ref. \\
    \hline
    $\Gamma_\mathrm{d}$ & $\left(3.21 \pm 1.55\right)\Gamma_\mathrm{t}$ & $2$ & App.~\ref{apx:photonflux}\\      
    $\gamma_\mathrm{S}$ & $\left(1.69 \pm 1.00\right)\overline{\Delta_\mathrm{S}}$ & $1$ & Eq.~\ref{eq:dephasing} \\ 
    $\sigma_\mathrm{g}$ & $\left(0.04 \pm 0.04\right)\sigma_\mathrm{geo}$ & $0.06$ & \citep{vitrant1982rydberg} \\     
    $\sigma_\mathrm{e}$ & $\left(1.10 \pm 0.55\right)\sigma_\mathrm{geo}$ & \SIrange{1}{10}{} & \citep{nagesha2003electron}   
    \end{tabular}
    \caption{Fitted parameter with 95\% confidence intervals from the non-linear least-squares optimization,
    and expected reference values.}
    \label{tab:params}
\end{table}
The data 
covered a wide range of different experimental parameters
as shown in Fig.~\ref{fig:simulation:measurement}.
We excluded the values close to the plasma formation edge
$\mathrm{\Delta}_R > 0$, for the dashed traces.
The steep waveform has a disproportionate contribution to the overall error measure,
and the point of plasma formation is prone to additional uncertainties,
such as fluctuations in charge density due to the cell walls which can trigger the formation of the plasma state.
Furthermore, as previously shown in Fig.~\ref{fig:universal},
the edge frequency for the plasma formation point
depends on the raw Rabi frequency,
while the line shape of the system
is defined by the \emph{adjusted} Rabi frequency.
The results of the fits shown in Tab.~\ref{tab:params} demonstrate that our assumptions are sound,
as all the fit values are close to the expected values. 
Perhaps most significantly, $\gamma_\mathrm{S}$
indicates that our approximation of the line broadening due to the Stark shifts of the plasma
is in good agreement with the proposed mechanism. Likewise, $\Gamma_\mathrm{d}$ is consistent with our estimates using the photon emission rate suggesting that our modelling of the recombination rate is consistent with the experimental observations.  
The cross-section $\sigma_\mathrm{g}$,
which accounts for the finite ionization efficiency of Rydberg-ground state collisions
as compared to the geometric cross-section $\sigma_\mathrm{geo}$,
is in agreement with prior experiments
\citep{vitrant1982rydberg}, $0.06 \sigma_\mathrm{geo}$  
and similar to the very crude estimate in \citep{weller2016charge},
$0.03 \sigma_\mathrm{geo}$.
The ionization cross-section of the electron-Rydberg collisions,
$\sigma_\mathrm{e}$,
also yields a reasonable value.
The coulombic interaction in conjunction with the large polarizability of the Rydberg atom
gives an expected cross-section that that is larger than the geometric cross-section
\citep{nagesha2003electron,deutsch2005calculated}.
Calculating cross-sections for the electron-impact ionization of Rydberg states is beyond the scope of this manuscript,
but $\sigma_\mathrm{e}$ in Tab.~\ref{tab:params} is consistent
since it is slightly larger than $\sigma_\mathrm{geo}$.
Overall, the values that were fit to get best agreement with the spectra are consistent with expected results.
From the overall agreement, we conclude that the model is consistent with the spectra.
Figure~\ref{fig:simulation:measurement} shows the bistable fluorescence spectra,
as captured by the PMT when scanning the 1064-nm laser across the Rydberg resonance line.
Because the underlying mechanism is interactions between charged particles and Rydberg atoms,
the peak is shifted towards blue detuning as the charge density increases;
i.e., more atoms, or higher Rydberg fraction.
The shift direction is in agreement with the negative sign of the polarizability
one obtains for the $42\mathrm{D}_{5/2}$ state \cite{vsibalic2017arc}.
An increase in fluorescence amplitude is also observed as the ground state density,
and consequently the Rydberg density, increases.
The measurements in Fig.~\ref{fig:simulation:measurement}
are overlaid with the results from the model,
adjusted only for amplitude.
The solid and dashed lines correspond to positive, red to blue,
and negative, blue to red, scan directions, respectively.
We achieve excellent agreement between the calculations and the measurements
as demonstrated by the low residuals shown in the figure.
The spectral broadening and line shape are described well and
the hysteresis feature is correctly reproduced
for both red and blue 1064-nm detuning sweeps.
For certain settings a second hysteresis feature
on the seemingly \emph{wrong} side of the spectrum appears,
which has also previously been observed \citep{ding2016non,weller2016charge},
but is straightforward to explain.
For a blue shifting interaction,
one expects an effect on the blue detuned side of the resonance. 
If the spectral broadening is larger than the shift caused by the Stark effect,
the far-detuned features of the resonance can extend beyond the resonance position,
and influence the spectrum on the red detuned side of the overall resonance feature.
The Stark map of the 42D$_{5/2}$ state, Fig.~\ref{fig:starkmap},
features a red shifting branch,
and also the 42D$_{3/2}$ state is in close proximity.
However, both these aspects are neglected in the modeled traces,
but the feature is still reproduced correctly.
Despite the overall excellent agreement between the model and experiment,
something can be learned from the regions where the deviations are maximum.
For one, the plasma formation point is not always precisely predicted
by the simulation for the larger scaling parameters,
but the jump occurs earlier than expected.
We attribute the deviation to the fact that when the system is in the bistable regime,
minor disturbances,
e.g.,~laser fluctuations 
or random charges produced by the photoelectric effect on the vapor cell walls,
can cause the system to jump to the plasma state before reaching a detuning
where the jump would take place in the absence of perturbation.
Once the system is in the high population state,
it will stay there, even if the average parameter values
would not support the jump to the plasma state at this point.

\section{Conclusion}
\label{sec:conclusion}
We have shown measurements and a comprehensive numerical model
that describes optical bistability in a thermal Rydberg gas.
We showed that plasma formation is the key ingredient
in the phenomena of optical bistability in thermal Rydberg gases.
By measuring radio-frequency transmission spectra
we established that a plasma exists in the gas.
We characterized the plasma by direct measurement of the plasma frequency
as weakly ionized.
Measuring the fluorescence at $\Delta_\mathrm{E}$
and its dependence on detuning and 1064-nm laser sweep direction 
allows us to determine the plasma formation point. 
We combined all experimental parameters into scaling parameters $S$ and $S^*$,
such that the plasma formation point scales linearly with these parameters.
The excitation dynamics were characterized using the results.
Measuring the fluorescence at different positions along the cell demonstrates the role of absorption of the 455-nm laser.
We observed a \emph{lightsaber}-like fluorescing beam that extends only along a part of the cell length,
because the necessary conditions for the ionization avalanche can only be sustained when the laser is strong enough.
The fluorescence spectrum shows a broad range of Rydberg states
consistent with recombination from the plasma continuum.
We developed a numerical model based on 
collisional cross-sections
and motional dephasing in an electric field distribution due to the presence of the plasma,
that fits the spectral data extremely well confirming our microscopic model.
The physical parameters obtained from fits to the spectral data are in excellent agreement with expectations.
We conclude that plasma formation is the fundamental ingredient in optical bistability in thermal Rydberg gases.

\begin{acknowledgments}
The authors acknowledge contributions from N.~Sieber in an early stage of the experiment.
We thank S.~Weber for stimulating discussions.
J.P.S thanks the AFOSR (Grant No. FA9550-12-1-0282)
and NSF (Grant No. PHY-1104424) for support.
\end{acknowledgments}

\bibliography{references}
\appendix
\section{Adjusted Rabi frequencies}
\label{apx:absorption}
The laser beams are being absorbed
while propagating through the atomic medium.
Therefore, the Rabi frequency in general is a function of the position $z$,
$\Omega = \Omega\left(z\right)$,
c.f., Fig.~\ref{fig:cell}.
In our experiment,
the 1064-nm laser intensity is not considerably affected 
by absorption along the cell.
The blue 455-nm laser beam however is strongly absorbed, 
since it couples the highly populated ground state to the intermediate state.
As the \emph{local} $\Omega_\mathrm{B}$ varies along the length of the vapor cell,
the excitation and ionization conditions vary.
The spectrum one obtains by measuring the fluorescence therefore depends on the spatial position
at which the fluorescence is collected.
As a consequence,
one needs to adjust $\Omega_\mathrm{B}$
according to the position where the measurement is made,
which is indicated by $\Omega^\star$.
We numerically propagate the intensity through a thermal ensemble of the respective density
to determine $\Omega^\star$,
following the derivation in \citep{siddons2014light}.

\section{Lindblad Master Equation}
\label{apx:master}
The density matrix is given by
\begin{equation}
	\rho = 
    	\left(\begin{array}{cccc}
			\rho_\mathrm{g} & \rho_{1,2} & \rho_{1,3} & 0 \\
			\rho_{2,1} & \rho_{2,2} & \rho_{2,3} & 0 \\
			\rho_{3,1} & \rho_{3,2} & \rho_\mathrm{Ryd} & 0 \\
			0 & 0 & 0 & \rho_\mathrm{ion}
		\end{array}\right) ,
\end{equation}
where the states are coupled by the two lasers according to the Hamiltonian
\begin{equation}
	H = \hbar
	\left(\begin{array}{cccc}
		0  & \frac{\Omega_\mathrm{B}}{2} & 0 & 0 \\
		\frac{\Omega_\mathrm{B}}{2} & -\Delta_1 & \frac{\Omega_\mathrm{R}}{2} & 0 \\
		0 & \frac{\Omega_\mathrm{R}}{2} & -\Delta_1-\left(\Delta_2+\Delta_\mathrm{S}\right) & 0 \\
		0 & 0 & 0 & 0
	\end{array}\right)
\end{equation}
with Rabi frequencies $\Omega_\mathrm{B}$ and $\Omega_\mathrm{R}$
and the reduced Planck constant $\hbar$.
The detunings are $\Delta_1 = \Vec{k}_\mathrm{B}\cdot\Vec{v}$
and $\Delta_2 = \Delta_\mathrm{R} + \Vec{k}_\mathrm{R}\cdot\Vec{v}$
with wave vectors $\Vec{k}_\mathrm{B}$, $\Vec{k}_\mathrm{R}$ and
atom velocity $\Vec{v}$.
The detuning $\Delta_2$ includes the laser detuning $\Delta_\mathrm{R}$,
and the detuning resulting from the Doppler effect.
The 455-nm laser is assumed to be tuned to resonance.
Figure~\ref{fig:schematic}(a) shows the relevant near-resonant atomic levels.
In the case of the Rydberg state,
the additional term $\Delta_\mathrm{S}$ represents the Stark shift due to nearby charges,
Fig.~\ref{fig:schematic}(c).
The ground state and intermediate state Stark shifts are negligible.
The Lindblad operator, $L = L_{1}+L_{2}+L_{3}$ accounts for dephasing, decay and ionization mechanisms.
First, $L_{1}$,
the intermediate state decoherence and decay,
which is dominated by its natural lifetime
$\Gamma_{2,1}/2\pi = \SI{1.18}{\mega\hertz}$ \citep{schmieder1970level},
gives the contribution
\begin{equation}
	L_{1} = 
	\left(\begin{array}{cccc}
		\Gamma_{2,1} \rho_{2,2} & -\frac{1}{2}\Gamma_{2,1} \rho_{1,2} & 0 & 0 \\
		-\frac{1}{2}\Gamma_{2,1} \rho_{2,1} & -\Gamma_{2,1}\rho_{2,2} & -\Gamma_{2,1}\rho_{2,3} & 0 \\
		0 & -\frac{1}{2}\Gamma_{2,1} \rho_{3,2} & 0 & 0 \\
		0 & 0 & 0 & 0
	\end{array}\right) .
\end{equation}
Second, $L_{2}$ takes into account
ionization and recombination,
\begin{equation}
\setlength{\arraycolsep}{2pt}
\medmuskip = 1mu
	L_{2} = 
	\left(\begin{array}{cccc}
		\Gamma_\mathrm{d} \rho_\mathrm{ion} & 0 & -\frac{1}{2} \Gamma_\mathrm{i} \rho_{1,3} & 0 \\
		0 & 0 &  -\frac{1}{2} \Gamma_\mathrm{i} \rho_{2,3} & 0 \\
		-\frac{1}{2} \Gamma_\mathrm{i} \rho_{3,1} & -\frac{1}{2} \Gamma_\mathrm{i} \rho_{3,2} & -\Gamma_\mathrm{i} \rho_\mathrm{Ryd} & 0 \\
		0 & 0 & 0 & \Gamma_\mathrm{i} \rho_\mathrm{Ryd} - \Gamma_\mathrm{d} \rho_\mathrm{ion}
	\end{array}\right).
\end{equation}
The ionization rates are given by $\Gamma_\mathrm{i}$ while
the loss of charged particles due to recombination and motion out of the interaction region
are represented by $\Gamma_\mathrm{d}$.
The interaction between the Rydberg atoms and the plasma is the last part of $L$ and is modeled as
\begin{equation}
	L_{3} = 
	\left(\begin{array}{cccc}
		\Gamma_\mathrm{Ryd} \rho_\mathrm{Ryd} & 0 & -\frac{1}{2} \gamma \rho_{1,3} & 0 \\
		0 & 0 &  -\frac{1}{2} \gamma \rho_{2,3} & 0 \\
		-\frac{1}{2} \gamma \rho_{3,1} & -\frac{1}{2} \gamma \rho_{3,2} & -\Gamma_\mathrm{Ryd} \rho_\mathrm{Ryd} & 0\\
		0 & 0 & 0 & 0
	\end{array}\right) .
\end{equation}
The rate $\Gamma_\mathrm{Ryd}$ includes the natural lifetime of the Rydberg state
and transit time effects.
The dominating part here clearly is the transit time decay $\Gamma_\mathrm{t}$,
therefore we estimate $\Gamma_\mathrm{Ryd}/2\pi = \SI{0.2}{\mega\hertz}$
for the beam sizes used in the experiments.
The dephasing factor $\gamma$ accounts for the line broadening mechanisms.
\section{Recombination Rate and Transit Time Loss Rate Estimation}
\label{apx:photonflux}
Given the experimental parameters,
one can
estimate
the photon flux
emitted by plasma recombination.
Here we show the quantitative analysis
for the experimental parameters as well as measured and simulated values
from the dataset in Fig.~\ref{fig:simulation:measurement}.
The atomic densities cover a range from 
$\SI[per-mode=reciprocal]{5e10}{\per\cubic\centi\meter}$
to
$\SI[per-mode=reciprocal]{1.8e12}{\per\cubic\centi\meter}$.

The PMT has a sensitivity of $\SI{1.89}{\volt\per\nano\watt}$ 
and yields a signal height of 0.1 to \SI{0.7}{\volt}.
Given the collection efficiency of the lens system capturing the fluorescence,
this corresponds to an emitted power of approximately 70 to \SI{360}{\nano\watt},
or a photon flux of
$\SI[per-mode=reciprocal]{2e11}{\per\second}$
to
$\SI[per-mode=reciprocal]{1.1e12}{\per\second}$,
when assuming only photons of $\SI{600}{\nano\meter}$ wavelength.
The simulation for these configurations,
c.f.~Sec.~\ref{sec:model},
peaks at a plasma population $\rho_\mathrm{ion}$ between
$\SI{0.8e-3}{}$
and
$\SI{9.6e-3}{}$.
A sphere with a diameter of \SI{1.5}{\milli\meter} 
-- approximately the volume we collect photons from --
therefore contains
$\SI{4.5e5}{}$ to
$\SI{4.8e6}{}$ ions.
Multiplied with the de-ionization rate reduced by transit time
(this gives the fluorescence photon emission rate),
we end up with a total number of 
\SI[per-mode=reciprocal]{2e11}{\per\second}
to
\SI[per-mode=reciprocal]{2.1e11}{\per\second},
which is in excellent agreement with the measured photon flux.
The simulated Rydberg population ranges between 
\SI[per-mode=reciprocal]{2.1e8}{\per\cubic\centi\meter}
and
\SI[per-mode=reciprocal]{2.5e8}{\per\cubic\centi\meter},
which is around \SI{4e5}{} Rydberg excitations per 1.5-mm sphere.
With regard to the transit time losses,
we estimate \mbox{$\Gamma_\mathrm{t}/2\pi = \SI{0.2}{\mega\hertz}$}
for the given beam sizes.
By multiplying the rate with the ion numbers per 1.5-mm sphere from above,
we obtain values for the transit time ion loss between
$\SI[per-mode=reciprocal]{9e10}{\per\second}$
and
$\SI[per-mode=reciprocal]{9.7e11}{\per\second}$,
roughly the same range as the photon flux.
\section{Electric field distribution}
\label{apx:efield}
\begin{figure}[htbp]
	\includegraphics[]{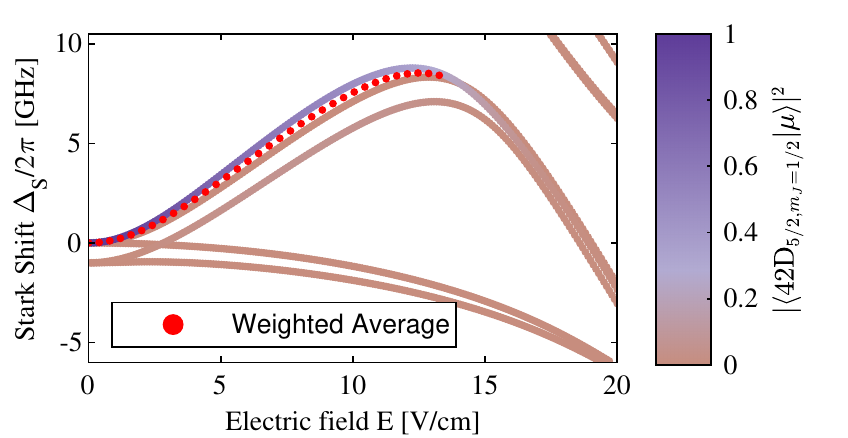}
    \caption{Stark map for the $42\mathrm{D}_{5/2}$ state
    with all 3 $m_\mathrm{J}$ sublevels,
    and approx.~\SI{1}{\giga\hertz} below the $42\mathrm{D}_{3/2}$ state.
    The overlap to the $42\mathrm{D}_{5/2}, m_\mathrm{J} = 1/2$
    is indicated by the color code
    for electric field in parallel to the laser polarization.
    To get an effective mapping from electric field strength to a certain Stark shift,
    we integrate over all angles for the electric field direction,
    and weight with the respective overlap to the state the lasers couple to.
    The dotted line indicates the mapping we used for the Stark shift in our model.}
    \label{fig:starkmap}
\end{figure}

The interesting part of the fluorescence spectrum for the $42\mathrm{D}_{5/2}$ Rydberg state
occurs on the blue detuned side of the Rydberg resonance,
where the plasma suddenly forms.
A Stark map for the $42\mathrm{D}_{5/2}$ Rydberg state,
calculated using \citep{weber2017calculation},
is shown in Fig.~\ref{fig:starkmap}.
The particular state has 3 sub-levels and splits up into two branches.
The dominant branch has a negative polarizability,
and consists of $m_\mathrm{J} = 1/2$ and $3/2$.
The second branch, $m_\mathrm{J} = 5/2$,
evokes a red detuning Stark shift.
Since the electric field caused by randomly positioned charges
has no preferred direction,
we integrate over all angles with respect to the laser polarization axis.
The dominant contribution to the Stark shift for each electric field strength
is then obtained via the angular integral,
weighted by the respective overlap
with the (unperturbed) 42D$_{5/2}$, $m_\mathrm{J} = 1/2$ state.
Since both positive and negative branches are well separated
for most of the electric field strengths,
we only consider the blue-shifting branch for our calculations.
The electric field distribution is modeled by randomly positioned point charges which
depends on the charge density $\mathcal{N}$.
The distribution, $\mathcal{P}_\mathcal{N}\left(E\right)$,
is described by the Holtsmark probability distribution function \citep{holtsmark1919verbreiterung}
\begin{equation}
  \mathcal{P}_\mathcal{N}\left(E\right) =
  \mathcal{H}\left(E/Q_\mathcal{H}\right)/Q_\mathcal{H} ,
  \label{eq:holtsmark}
\end{equation}
where the normal field is given by
\begin{equation}
  Q_\mathcal{H} = 
  \frac{e}{2\epsilon_0}
  \left(\frac{4}{15} \mathcal{N}\right)^\frac{2}{3} ,
\end{equation}
and
\begin{equation}
  \mathcal{H}\left(\beta\right) = 
  \frac{2}{\pi\beta}\int\limits_0^\infty \mathrm{d}x\,
  x \sin{\left(x\right)} \exp{\left(-\left(x/\beta\right)^{3/2}\right)} .
\end{equation}
The electric field strengths $E$ are converted into Rydberg energy level shifts
via the angular average of the Stark map for the Rydberg state, 
as calculated in the previous paragraph.
We used only the ion density for the electric field distribution calculations.
The electrons move much faster than ions,
and the energy shifts due to their electric fields averages out
on a timescale that is shorter than the atomic dynamics.
\section{Iterative Equilibrium}
\label{apx:iterative}
\begin{figure}[htbp]	
\begin{tikzpicture}[]
\tikzstyle{ann} = [draw=none, fill=none, right]
\matrix[nodes={thick, fill=none},row sep=1pt,column sep=1pt] {
	\node[draw, dashed, ellipse, minimum height=20pt] (A) at (0,1.6) {initial values};
	\node[draw, regular polygon, regular polygon sides = 5, align=center, yscale=0.35, minimum width=150pt] (B) at (0,-0.1) {};
	\node[align=center] (0,1) {\textbf{charge density}\\ electric field distribution};	
	\node[draw, ellipse, minimum height=25pt] (C) at (-2.25,-1.7) {broadening};
	\node[draw, ellipse, minimum height=25pt] (D) at (0,-1.7) {shift};
	\node[draw, ellipse, minimum height=25pt] (E) at (2.25,-1.7) {ionization};
    \node[draw, rectangle, minimum height=25pt, text width=180pt, align=center] (F) at (0,-2.9) {\textbf{Lindblad Master Equation}\\ steady state solution};
	\node[draw, ellipse, minimum height=25pt] (G) at (-3.7,-0.55) {averaging};
	\draw[thick,dashed,->, to path={-| (\tikztotarget)}]	(A) -- (B);
	\draw[thick,-, to path={-| (\tikztotarget)}]	(B) -- (C);
	\draw[thick,-, to path={-| (\tikztotarget)}]	(B) -- (D);
	\draw[thick,-, to path={-| (\tikztotarget)}]	(B) -- (E);
	\draw[thick,->, to path={-| (\tikztotarget)}]	(C.south) -- (C.south|-F.north);
	\draw[thick,->, to path={-| (\tikztotarget)}]	(D.south) -- (D.south|-F.north);
	\draw[thick,->, to path={-| (\tikztotarget)}]	(E.south) -- (E.south|-F.north);
	\draw[thick,->, to path={-| (\tikztotarget)}]	(F) edge (G);
	\draw[thick,->, to path={-| (\tikztotarget)}]	(G.north) |- +(1.19,0.28);
	\node[ann]{};\\
};
\end{tikzpicture}
    \caption{Flow chart of the algorithm. 
    At each iteration,
    the current charge density determines an electric field distribution
    and hence the broadening and shift of the Rydberg energy level.
    The charge density also influences the ionization rate.
    With these values, the ensemble of steady-state solutions of the master equation is calculated.
    When averaged over all velocity and electric field components, 
    an improved estimation of the charge density at each laser detuning is achieved.
    The procedure is repeated until the system converges to a steady-state solution.
    }
    \label{fig:algorithm}
\end{figure}
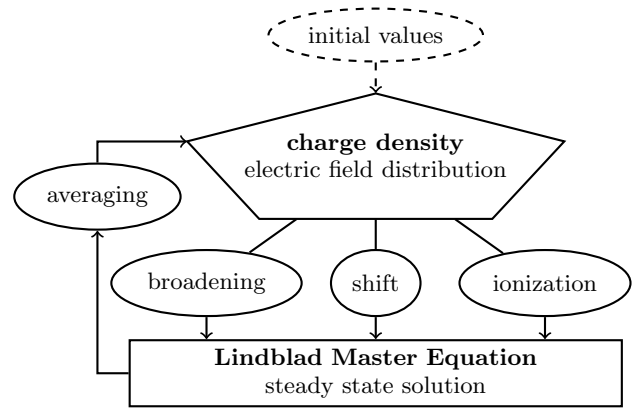

For a given initial charge density, $\Omega_\mathrm{B}$, $\Omega_\mathrm{R}$,
atomic density, temperature and Stark shift distribution,
a steady state solution of the master equation is obtained
that yields a charge density
\begin{equation}
	\mathcal{N}' = \rho_\mathrm{ion}\mathcal{N}_\mathrm{g} .
\end{equation}
The calculated charge density is used for a subsequent calculation.
The equations are solved iteratively
using the previous calculation of the charge density
as an input parameter for subsequent computation steps
until the system converges to an equilibrium solution.
A flow-chart summarizing this algorithm is shown in Fig.~\ref{fig:algorithm}.
\end{document}